\documentclass[aps,ams,prl,twocolumn,groupedaddress,showpacs]{revtex4}

\usepackage{graphicx}
\usepackage{dcolumn}
\usepackage{bm}

\newcommand{\ket}[1]{| #1 \rangle}

\begin{document}

\title{Uncollapsing of a quantum state in a superconducting phase qubit}

\author{Nadav Katz,$^{1}$\footnote{Current address: The Racah
Institute of Physics, The Hebrew University, Jerusalem, 91904,
Israel} Matthew Neeley,$^{1}$ M. Ansmann,$^{1}$ Radoslaw C.
Bialczak,$^{1}$ M. Hofheinz,$^{1}$ Erik Lucero,$^{1}$ A.
O'Connell,$^{1}$ H. Wang,$^{1}$ A. N. Cleland,$^{1}$ and John M.
Martinis,$^{1}$   and Alexander N. Korotkov$^{2}$}
\affiliation{
$^1$Department of Physics, University of California, Santa Barbara, CA 93106, USA \\
$^2$Department of Electrical Engineering, University of California,
Riverside, CA 92521, USA}

\date{\today}

\begin{abstract}
We demonstrate in a superconducting qubit the conditional recovery
(``uncollapsing'') of a quantum state after a partial-collapse
measurement. A weak measurement extracts information and results in
a non-unitary transformation of the qubit state. However, by adding
a rotation and a second partial measurement with the same strength,
we erase the extracted information, effectively canceling the effect
of both measurements.  The fidelity of the state recovery is
measured using quantum process tomography and found to be above 70\%
for partial-collapse strength less than 0.6.
\end{abstract}


\maketitle

The observation of a quantum system necessarily perturbs the state of
the system. For a strong measurement, the quantum state is understood
to collapse irrevocably to one of the eigenstates of the measurement
operator; this concept of projective measurement is a central paradigm
of modern physics \cite{Wheeler-book}. For weak measurements,
however, the collapse is now understood to be partial, with correspondingly
partial information drawn from the measurement yielding a non-unitary
transformation of the quantum state.  It has been predicted that
after such a weak measurement, the initial quantum state of the system
can be recovered by essentially undoing the effect of the measurement
\cite{Ueda, QUDProposalPRL} and causing a quantum ``uncollapsing''.

Superconducting phase qubits provide an excellent system for testing
this concept of ``uncollapsing''. Our experimental implementation
\cite{MartinisFirst} uses a controlled measurement process whose
projective strength can be tuned continuously from a weak partial
measurement to a full projective one
\cite{PartialExperimentScience}. Using this system, we can
experimentally test reversing the partial, measurement-induced
collapse of a quantum state. Similar tests of partial or continuous
weak measurements should also be possible for other types of
solid-state qubits \cite{s-s-qubits}.

\begin{figure} [pb]
\includegraphics[width=8.5cm]{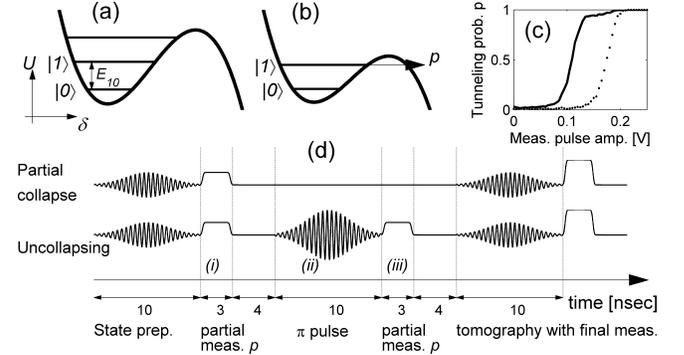}
\caption{ (a) The qubit potential during application of the
microwave pulses, which cause coherent transitions between states
$\ket{0}$ and $\ket{1}$. The phase of the Josephson junction is
$\delta$. (b) During a partial measurement the state $\ket{1}$
tunnels out of the well with probability $p$. (c) The tunneling
probability $p$ (solid line) is determined by the amplitude of the
measurement pulse which lowers the barrier; we use sufficiently
small amplitude to avoid tunneling from the state $\ket{0}$ (dotted
line). The maximal measurement visibility (the difference between
the solid and dotted lines) is about $90\%$. (d) Pulse sequences
(including state tomography) for the partial-collapse experiment
(upper trace, as in \protect\cite{PartialExperimentScience}) and for
the quantum uncollapsing (lower trace). The effect of the partial
measurement [step (i)] is undone by applying the $\pi$-pulse [step
(ii)] and additional (uncollapsing) partial measurement [step (iii)]
with the same strength $p$.} \label{Fig:PulseSequences}\end{figure}

In our experiment, the superconducting phase qubit is prepared in a
combination of its ground $\ket{0}$ and first excited $\ket{1}$
states. A partial measurement of the qubit then yields a
``detection'' event, which occurs with probability $p$ when the
qubit is in the $\ket{1}$ state, while it never occurs for the qubit
in the $\ket{0}$ state. If the measurement yields a null result
(i.e. no event detected), this leads to the partial collapse of the
qubit state towards $\ket{0}$. This evolution towards the $\ket{0}$
state is driven by the extracted information, and does not involve
any energy exchange. We then employ the following method (proposed
by Jordan and one of the authors \cite{QUDProposalPRL}) to
``uncollapse'' the result of the measurement [see pulse sequence in
Fig.\ \ref{Fig:PulseSequences}(d)]: After the preparation of an
arbitrary initial state of the qubit and (i) partial collapse due to
null-result measurement with strength $p$, we (ii) apply a
$\pi$-pulse, coherently swapping the amplitudes of the qubit states
$\ket{0}$ and $\ket{1}$, and (iii) partially measure the qubit state
again, with the same measurement strength $p$ \cite{DropPI}. The
combination of steps (ii) and (iii) ``anti-symmetrizes'' the
information extracted from the first measurement of the qubit, and
with an overall probability $1-p$ of two null results \cite{probab},
regardless of the initial state, the qubit state is coherently
restored to its initial, pre-measurement state (here including a
$\pi$-rotation \cite{DropPI,QuantumEfficiency}).

In order for the uncollapsing procedure to work, we have to erase the
information that was already extracted classically. This distinguishes
this measurement-induced uncollapsing from a ``quantum eraser''
\cite{Scully-eraser}, in which only potentially extractable information
is erased. Note also that the result of the second measurement is stochastic,
and only a particular result will succeed in erasing the information,
thus leading to a less than unity probability of success for uncollapsing;
however, in the case of this desired result, the qubit's initial state
is fully recovered.

In this Letter, we report the first experimental demonstration
of quantum uncollapsing by implementing the above described protocol, where we obtain fidelities well over 70\%, quantified by quantum process
tomography \cite{IkeAndMike}. Besides confirming the ability to undo a
partial quantum measurement, this result confirms the high quantum
efficiency of our measurement.

Our superconducting phase qubit has been described in detail
previously \cite{MartinisFirst}. We briefly review the relevant
details and modifications here. The qubit is fabricated as a
superconducting loop interrupted by a $\sim 1$ $\mu m^2$ sized
Josephson junction of critical current $2 \mu$A, shunted by a low
loss-tangent parallel plate capacitor ($1$pF) formed with a-Si:H
dielectric. We initialize the system in the ground state of the
qubit's cubic-shaped potential well [see Fig.\
\ref{Fig:PulseSequences}(a)]. The logic qubit is formed by the
ground state $\ket{0}$ and the first excited state $\ket{1}$ of this
well (separated by $E_{10}/h=6.75$ GHz, with $h$ Planck's constant).
A coherent initial state is prepared by a shaped (in both phase and
amplitude) microwave pulse with nanosecond time resolution and 14
bit precision. We use on-resonance  \cite{StrongPIshfit} 10 ns long,
$4$ ns FWHM, Slepian pulses \cite{Slepian,FidelityErik} to ensure
optimal spectral properties (minimizing unwanted excitation of
higher states of the well \cite{FidelityErik}) while avoiding pulse
overlap in the time domain. The resulting (initial) qubit state can
be written as
   \begin{equation}
\ket{\psi_0}=\cos(\theta_0/2)\ket{0}+e^{-i\phi_0}\sin(\theta_0/2)\ket{1}.
    \label{init}\end{equation}

The partial measurement [step (i) of the protocol] is done in the
same way as in Ref.\ \cite{PartialExperimentScience}. By applying a
short (3 ns) bias pulse, we lower the quantum well barrier [Fig.\
\ref{Fig:PulseSequences}(b)] that leads to the selective tunneling of
the $\ket{1}$ state out of the well. The probability $p$ for this
tunneling to occur (i.e. the measurement strength) can be tuned continuously
from 0 to 1 by varying the bias pulse amplitude [Fig.\
\ref{Fig:PulseSequences}(c)]. The tunneling event is registered at a
later time with an on-chip SQUID that easily distinguishes between
states remaining in the qubit well and those that tunneled out.
    For the initial state given by Eq. (\ref{init}), the tunneling occurs with
probability $p\,\sin^2(\theta_0/2)$. If no tunneling occurs (null
result), the initial state $\ket{\psi_0}$ changes (partially collapses) to
    \begin{eqnarray}
&&
\ket{\psi_M}=\cos(\theta_M/2)\ket{0}+e^{-i(\phi_0+\phi_M)}\sin(\theta_M/2)\ket{1},
\quad
    \label{collapse-1}    \\
&& \theta_M=2\tan^{-1} [\sqrt{1-p}\, \tan(\theta_0/2)],
    \label{collapse-2}  \end{eqnarray}
where $\phi_M$ is an accumulated phase due to an adiabatic change in
the energy level spacing during the measurement. This information-related non-unitary transformation (confirmed in the
experiment \cite{PartialExperimentScience}) is precise
only in the ideal case. It neglects energy and phase
relaxation within the qubit well, which is an acceptable
approximation since the corresponding relaxation times $T_1=450$ ns
and $T^{*}_2=350$ ns (and $T_2=120$ ns) are significantly longer than the
experiment duration \cite{echoing}. Eqs.\ (\ref{collapse-1})--(\ref{collapse-2})
also neglect incoherence and noise in the process of virtual
tunneling; however, the theoretical analysis \cite{Pryadko} of the
density matrix evolution confirms that the simple result
(\ref{collapse-1})--(\ref{collapse-2}) is a good approximation.

\begin{figure}[t]
\centering
\includegraphics[width=8cm]{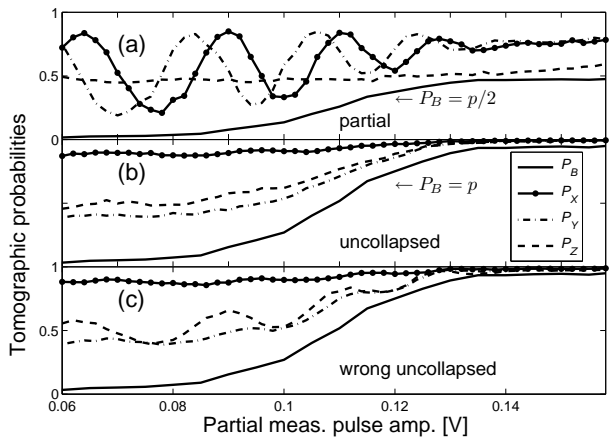}
\caption{The qubit tunneling probabilities $P_X$, $P_Y$, and $P_Z$
after the partial and tomographic ($X,Y,Z$) measurements for (a) the
partial-collapse sequence, (b) the uncollapsing sequence, and (c) a
``wrong'' uncollapsing with $\pi$-pulse replaced by $0.9\pi$-pulse.
Initial state is $(\ket{0}+\ket{1})/\sqrt{2}$. The background $P_B$
is the probability of qubit tunneling before the state tomography
(see text).
 }
\label{xyzResults}\end{figure}

\begin{figure}[t]
\centering
\includegraphics[width=8.5cm]{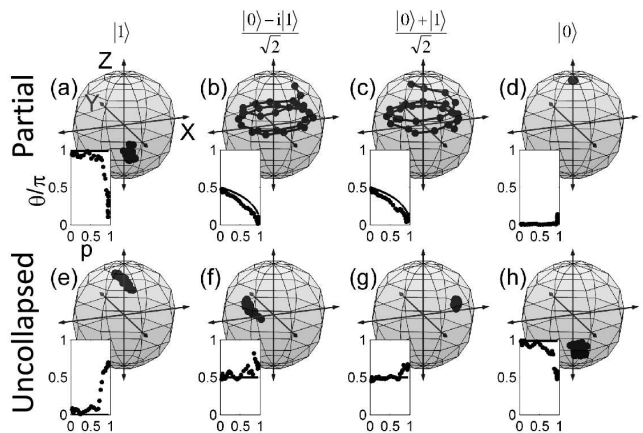}
\caption{The qubit states on the Bloch sphere, as measured by the
state tomography, after the partial collapse (first row) and
uncollapsing (second row). Initial state is $\ket{1}$ for panels (a)
and (e), $(\ket{0}-i\ket{1})/\sqrt{2}$ for (b) and (f),
$(\ket{0}+\ket{1})/\sqrt{2}$ for (c) and (g), and $\ket{0}$ for (d)
and (h). The points shown on the Bloch spheres (connected by lines as a guide for eye)
correspond to varied measurement strength up to $p=0.7$. For ideal
uncollapsing the states in the second row should not depend on $p$.
The insets in each panel show the measured (dots) and theoretical
(line) dependence of the polar angle $\theta /\pi$ on $p$.
 }
\label{BlochSpheres}\end{figure}

    The qubit state after the partial collapse is analyzed by state tomography (as in
\cite{PartialExperimentScience,TomographyMatthiasPRL,TomographyYale,TomographyProposalNori}),
consisting of 3 types of tomographic rotations (either a $\pi/2$-pulse
rotating about the $Y$-axis of the Bloch sphere, a $\pi/2$-pulse
rotating about the $X$-axis, or no rotation) followed by a full measurement (with
$p\approx 1$) -- see the upper trace in Fig.\
\ref{Fig:PulseSequences}(d). In this way we measure the qubit
tunneling probabilities $P_X$, $P_Y$, and $P_Z$, which correspond to
the qubit state components $X$, $Y$, and $Z$ on the Bloch sphere (in
the rotating frame). Since $P_X$, $P_Y$, and $P_Z$ include both the
probability of tunneling during the tomography measurement and the
background probability $P_{B}=p \, \sin^2 (\theta_0/2)$ due to the
partial measurement, the qubit state components are given by
$\{X,-Y,-Z\}=2(P_{\{X,Y,Z\}}-P_B)/(1-P_B)-1$ (the minus signs on $Y$ and $Z$ come from following the
convention setting the $\ket{0}$ at $Z=+1$). The measured
tunneling probabilities $P_X$, $P_Y$, and $P_Z$ for the initial
state $(\ket{0}+\ket{1})/\sqrt{2}$ are shown in Fig.\
\ref{xyzResults}(a), as functions of the pulse amplitude for partial
measurement (which is in a one-to-one correspondence with $p$); in
this case $P_B=p/2$ [also shown in Fig.\ \ref{xyzResults}(a)]. Note
the large oscillations in $P_X$ and $P_Y$, indicating that the
partial measurement is accumulating a significant phase $\phi_M$, as
was seen in \cite{PartialExperimentScience}. The qubit state
components $X$, $Y$, and $Z$ calculated from the data in Fig.\
\ref{xyzResults}(a) are shown on the Bloch sphere in Fig.\
\ref{BlochSpheres}(c).

In order to recover the initial quantum state, we now add steps (ii)
and (iii) of the uncollapsing protocol -- see the lower trace in
Fig.\ \ref{Fig:PulseSequences}(d). The $\pi$-pulse about the $X$-axis
[step (ii)] after the partial collapse exchanges the basis states in
Eq.\ (\ref{collapse-1}), creating the qubit state
$\ket{\psi_\pi}=\sin(\theta_M/2)\ket{0}+e^{i(\phi_0+\phi_M)}
\cos(\theta_M/2)\ket{1}$. The second partial measurement with the
same strength $p$ [step (iii)] can either result in a tunneling
event, or not. In the case of no tunneling (null result again) the
partial-collapse evolution $\ket{\psi_\pi}\rightarrow \ket{\psi_F}$
is described by the same transformation as $\ket{\psi_0}\rightarrow
\ket{\psi_M}$ [see Eqs.\ (\ref{init})--(\ref{collapse-2})], and
therefore produces the state
$\ket{\psi_F}=\sin(\theta_0/2)\ket{0}+e^{i\phi_0}
\cos(\theta_0/2)\ket{1}$. As expected, $\ket{\psi_F}$ coincides with
the initial state $\ket{\psi_0}$ up to a $\pi$-rotation about
the $X$-axis. Notice that not only the polar angle $\theta_0$ is
restored (which is essentially the uncollapsing), but the azimuth
angle shift $\phi_M$ is also canceled (due to the usual spin echo
effect).

    The state tomography of the uncollapsed state $\ket{\psi_F}$ is
done in the same way as for the partially collapsed state
$\ket{\psi_M}$. The only difference is that now the background
probability $P_B$ is due to both partial measurements, and therefore
$P_B=1-[1-p\sin^2(\theta_0/2)][1-p\cos^2(\theta_M/2)]=p$,
independent of the initial state. The measured probabilities $P_X$,
$P_Y$, and $P_Z$ for the initial state $(\ket{0}+\ket{1})/\sqrt{2}$
are shown in Fig.\ \ref{xyzResults}(b) as functions of the
measurement pulse amplitude, and the corresponding qubit states on
the Bloch sphere are shown in Fig.\ \ref{BlochSpheres}(g). Notice
that compared to the partial-collapse results, the oscillations in $P_X$ and $P_Y$
are clearly suppressed (spin echo) and the qubit state
is restored to the equatorial plane. The measured state is quite
close to the ideal result of uncollapsing
$(\ket{0}+\ket{1})/\sqrt{2}$ for $p\leq 0.6$ (see below).
    If we purposefully change the $\pi$ pulse of the step (ii) to
a $0.9\pi$ pulse, the oscillations of $P_Y$ and $P_Z$ are somewhat
recovered [see Fig.\ \ref{xyzResults}(c)], and the qubit state moves
significantly out of the equatorial plane (not shown), indicating
that the uncollapsing procedure performance is degraded.

    So far we discussed experimental uncollapsing of the initial
state $(\ket{0}+\ket{1})/\sqrt{2}$. However, {\it any} initial state
should be restored by the same procedure. Instead of examining all
initial states to check this fact, it is sufficient to choose 4
initial states with linearly independent density matrices and use
the linearity of quantum operations \cite{IkeAndMike}. We choose initial
states $\ket{1}$, $(\ket{0}-i\ket{1})/\sqrt{2}$,
$(\ket{0}+\ket{1})/\sqrt{2}$ and $\ket{0}$. The corresponding qubit
states on the Bloch sphere after the partial collapse and after
uncollapsing are shown in Fig.\ \ref{BlochSpheres} for measurements
with a range of measurement strength $p$. For clarity we only show the results for
$p\leq 0.7$, since beyond this range our simple theory becomes too
inaccurate.
    The main reason why the protocol begins to fail for large $p$
is a noticeable probability $p_r\sim 0.1$ of energy relaxation to
the ground state during our 44 ns long sequence ($T_1 =450$ ns).
The relative contribution of such cases increases with $p$ and becomes very significant when the
selection probability $1-p$ becomes comparable to $p_r$, thus
ruining the fidelity of uncollapsing. Also notice that we use the
experimentally determined $p$, as shown in
\ref{Fig:PulseSequences}(c), which contains an approximate $5\%$
error due to state preparation and measurement
infidelities. The data is not re-scaled to correct for this error.

    Uncollapsing of the states $\ket{0}$ and $\ket{1}$ is straightforward
(they do not change in null-result measurements), so the small
deviations from the ideal results on the Bloch spheres in the left
and right columns of Fig.\ \ref{BlochSpheres} characterize the
imperfections of our experiment. Uncollapsing of the states
$(\ket{0}-i\ket{1})/\sqrt{2}$ and $(\ket{0}+\ket{1})/\sqrt{2}$ is
non-trivial; however, we see that the small deviations in Figs.\
\ref{BlochSpheres}(f) and \ref{BlochSpheres}(g) from the theoretical
result are approximately the same as for the trivial cases, thus
indicating that the uncollapsing itself is nearly ideal. Besides the
Bloch spheres, in Figs.\ \ref{BlochSpheres}(a-h) we also show the
dependence of the corresponding polar angles $\theta$ on the
measurement strength $p$. The small discrepancy with the theory
(with no fit parameters) is mainly due to intrinsic decoherence of
the qubit and measurement error. As discussed above, the discrepancy
becomes significant when $p$ approaches 1.

\begin{figure}[t]
\vspace{0.3cm}\centering
\includegraphics[width=8cm]{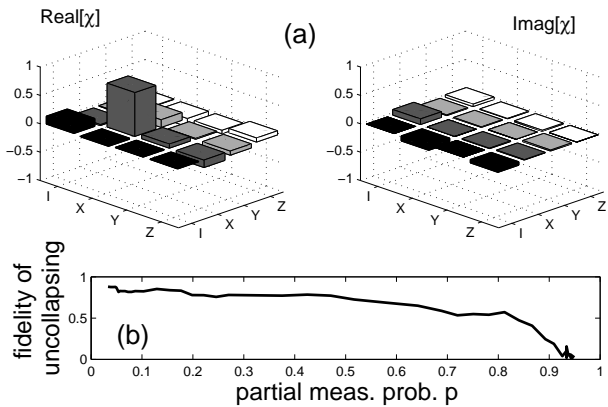}
\caption{(a) The quantum process tomography matrix $\chi$ for the
uncollapsing with $p=0.47$. (b) The fidelity of the quantum uncollapsing as a
function of the partial measurement probability $p$ \cite{errors}.
 }
\label{QPTResults}\end{figure}

    The state tomography for these four initial states is
sufficient for full characterization of the quantum process
tomography (QPT)
\cite{IkeAndMike,QPTIons,TLSMemoryMatthew,SQISRadek}. In Fig.\
\ref{QPTResults}(a) we show the QPT matrix $\chi$ in the standard
Pauli-matrix basis ($I$, $\sigma_X$, $\sigma_Y$, $\sigma_Z$),
generated by applying the conventional linear algebra formalism
\cite{IkeAndMike} to our results \cite{QPTprotocol} for the
uncollapsing protocol with $p=0.47$. As expected, we see a clear
peak at the $(X,X)$ location, indicating that the process is mainly
that of a $\pi$ rotation about the $X$ axis. The uncollapsing
fidelity is defined as the overlap of the $\chi$-matrix with the
ideal one (of a perfect $\pi$ pulse), i.e. the fidelity is simply
$\mbox{Re}\, \chi(X,X)$. The dependence of this fidelity on the
measurement strength $p$ is presented in Fig.\ \ref{QPTResults}(b),
which shows that the uncollapsing fidelity remains above $70\%$
until the degradation of the state recovery at $p \agt 0.6$ (because
of the reason discussed above).

In conclusion, we demonstrate a conditional uncollapsing of a
partially measured quantum state, and quantify this process by
quantum process tomography. While our protocol has apparent
similarity with the spin echo sequence (and includes the azimuth
angle recovery due to the echo effect), we emphasize the clear
difference between the two effects: the spin echo is the undoing of
an unknown unitary transformation, while uncollapsing is the undoing
of a known but non-unitary transformation.

The authors thank A.\ Jordan for useful discussions.
 Devices were made at the UCSB and Cornell Nanofabrication
Facilities, a part of the NSF funded NNIN network. This work was
supported by NSA and IARPA under ARO grant W911NF-04-1-0204 and by
NSF under grant CCF-0507227.


\vspace{-0.2cm}

\begin{thebibliography}{99}

\vspace{-0.3cm}


\bibitem{Wheeler-book} {\it Quantum theory and measurement}, edited
by J. A. Wheeler and W. H. Zurek (Princeton Univ. Press, 1983).

\bibitem{Ueda} M. Koashi and M. Ueda, Phys. Rev. Lett. {\bf 82},
2598 (1999); M. A. Nielsen and C. M. Caves, Phys. Rev. A {\bf 55},
2547 (1997); H. Mabuchi and P. Zoller, Phys. Rev. Lett. {\bf 76},
3108 (1996).

\bibitem{QUDProposalPRL}A. N. Korotkov and A. N. Jordan, Phys. Rev.
Lett. {\bf 97}, 166805 (2006);
    A. N. Jordan and A. N. Korotkov, in
``{\it Coherence and Quantum Optics IX}'', edited by N. P. Bigelow
et al. (Optical Soc. of America, 2008), p. 191.


\bibitem{MartinisFirst} J. M. Martinis, S. Nam, J. Aumentado, C. Urbina,
    Phys. Rev. Lett. {\bf 89}, 117901 (2002).
    K. B. Cooper et al., Phys. Rev. Lett. {\bf 93}, 180401 (2004).

\bibitem{PartialExperimentScience} N. Katz et al., Science {\bf 312}, 1498
(2006).

\bibitem{s-s-qubits} T. Yamamoto et al., Nature {\bf 425}, 941 (2003); J. H.
Plantenberg et al., Nature {\bf 447}, 836 (2007); J. Majer et al.,
Nature {\bf 449}, 443 (2007); J. R. Petta et al., Science {\bf 309},
2180 (2005).

\bibitem{DropPI} The original proposal \cite{QUDProposalPRL} called for
another $\pi$ pulse to return to the original basis. Without loss of
generality, we forgo this pulse in order to shorten the overall
sequence.

\bibitem{probab} This probability coincides with the upper bound
      for the success probability obtained in the general theory of
      uncollapsing \protect\cite{QUDProposalPRL}. Obviously, the
      success probability for uncollapsing decreases with increasing
      strength $p$ of the partial measurement, and reaches zero for the projective
      collapse ($p=1$).

\bibitem{QuantumEfficiency} Erasing the information is only
a necessary condition for uncollapsing; we also have to use a
measurement with 100\% quantum efficiency (so that quantum
information does not leak to an unmeasurable environment) and
compensate for a possible unitary transformation.

\bibitem{Scully-eraser} M. O. Scully and K. Dr\"uhl, Phys. Rev. A {\bf
25}, 2208 (1982).

\bibitem{IkeAndMike} M. A. Nielsen, I. L. Chuang, {\it Quantum Computation
and Quantum Information} (Cambridge Univ. Press, Cambridge, 2000).


\bibitem{StrongPIshfit} We note that such short pulses lead to
power dependent phase shifts [as measured by Frederick W. Strauch et
al., IEEE Trans. Appl. Superconductivity {\bf 17}, 105 (2007)].
These shifts were compensated for by a +8 MHz detuning of the $\pi$
pulses, while no shifts were needed for the $\pi/2$ pulses.



\bibitem{Slepian} D. Slepian,  The Bell system technical journal {\bf 57}, 1371
(1978).

\bibitem{FidelityErik} Erik Lucero et al., arXiv:0802.0903 (2008).

\bibitem{echoing} Because of the similarity of the uncollapsing to the echo
sequence, it is $T^{*}_2$ that determines the timescale for decay due to dephasing.

\bibitem{Pryadko} L. P. Pryadko and A. N. Korotkov, Phys. Rev. B 76, 100503(R) (2007).

\bibitem{TomographyMatthiasPRL} M. Steffen et al., Phys. Rev. Lett. {\bf 97}, 050502
(2006).


\bibitem{TomographyYale} A. A. Houck et al.,  Nature {\bf 449}, 328
(2007).

\bibitem{TomographyProposalNori} Yu-xi Liu, L. F. Wei and Franco Nori,
Europhys. Lett., {\bf 67 (6)}, 874 (2004); Yu-xi Liu, L. F. Wei and
Franco Nori, Phys. Rev. B {\bf 72}, 014547 (2005).

\bibitem{QPTIons} M. Riebe et al., New J. Phys. {\bf 9} 211 (2007).

\bibitem{TLSMemoryMatthew} Matthew Neeley et al., Nature Phys., in press
(2008).

\bibitem{SQISRadek} Radoslaw C. Bialczak et al., Abstract: P33.00010, APS March Meeting
(2007).

\bibitem{QPTprotocol} We use the experimentally measured initial states
as input states for the process tomography (and do not assume the ideal states).

\bibitem{errors} For $p>0.9$ the error $p_r$ dominates the analysis, causing
the fidelity shown in Fig. \ref{QPTResults}(b), to become
unphysical.
\end{thebibliography}
\end{document}